\documentclass{article}

\usepackage[preprint,nonatbib]{neurips_2022}

\usepackage[utf8]{inputenc} % allow utf-8 input
\usepackage[T2A,T1]{fontenc}    % use 8-bit T1 fonts
\usepackage{hyperref}       % hyperlinks
\usepackage{url}            % simple URL typesetting
\usepackage{booktabs}       % professional-quality tables
\usepackage{amsfonts}       % blackboard math symbols
\usepackage{nicefrac}       % compact symbols for 1/2, etc.
\usepackage{microtype}      % microtypography
\usepackage{xcolor}         % colors
\usepackage{mathtools}      % equations
\usepackage{tikz}  
\usetikzlibrary{shapes,arrows,positioning}  
\usepackage[pdf]{pstricks}
\usepackage{bm}
\usepackage[russian,english]{babel}
\usepackage{enumitem}

\DeclareSymbolFont{cyrillic}{T2A}{cmr}{m}{}
\SetSymbolFont{cyrillic}{bold}{T2A}{cmr}{bx}{it}
\DeclareMathSymbol{\cya}{\mathord}{cyrillic}{224}
\DeclareMathSymbol{\cyp}{\mathord}{cyrillic}{208}
\DeclareMathSymbol{\cys}{\mathord}{cyrillic}{241}
\DeclareMathSymbol{\cyo}{\mathord}{cyrillic}{238}

\title{A Modular and Adaptive System for Business Email Compromise Detection}

\author{
Jan Brabec\textsuperscript{1,2}\thanks{Corresponding author: \texttt{janbrabe@cisco.com}}
\quad Filip \v{S}rajer\textsuperscript{1}
\quad Radek Starosta\textsuperscript{1}
\quad Tom\'{a}\v{s} Sixta\textsuperscript{1}
\quad Marc Dupont\thanks{Formerly at Cisco Systems}\\
\quad \textbf{Milo\v{s} Lenoch}\textsuperscript{1}
\quad \textbf{Ji\v{r}\'{i} Men\v{s}\'{i}k}\textsuperscript{1}
\quad \textbf{Florian Becker}\textsuperscript{1}
\quad \textbf{Jakub Boros}\textsuperscript{1}\\
\quad \textbf{Tom\'a\v{s} Pop}\textsuperscript{1,3}
\quad \textbf{Pavel Nov\'ak}\textsuperscript{1}
\\
\textsuperscript{1}Cisco Systems\\
\textsuperscript{2}Czech Technical University in Prague, Faculty of Electrical Engineering\\
\textsuperscript{3}Faculty of Mathematics and Physics, Charles University
}

\begin{document}

\maketitle

\begin{abstract}
   The growing sophistication of Business Email Compromise (BEC) and spear phishing attacks poses significant challenges to organizations worldwide. The techniques featured in traditional spam and phishing detection are insufficient due to the tailored nature of modern BEC attacks as they often blend in with the regular benign traffic. Recent advances in machine learning, particularly in Natural Language Understanding (NLU), offer a promising avenue for combating such attacks but in a practical system, due to limitations such as data availability, operational costs, verdict explainability requirements or a need to robustly evolve the system, it is essential to combine multiple approaches together. We present CAPE, a comprehensive and efficient system for BEC detection that has been proven in a production environment for a period of over two years. Rather than being a single model, CAPE is a system that combines independent ML models and algorithms detecting BEC-related behaviors across various email modalities such as text, images, metadata and the email's communication context. This decomposition makes CAPE's verdicts naturally explainable. In the paper, we describe the design principles and constraints behind its architecture, as well as the challenges of model design, evaluation and adapting the system continuously through a Bayesian approach that combines limited data with domain knowledge. Furthermore, we elaborate on several specific behavioral detectors, such as those based on Transformer neural architectures.
\end{abstract}

\section{Introduction}
As Business Email Compromise (BEC) and spear phishing attacks become increasingly sophisticated and widespread, they present significant challenges to organizations worldwide~\cite{ic3InternetCrime}.
These targeted attacks not only lead to enterprise compromise but also result in substantial monetary damage.
Traditional techniques, such as Bayesian spam filtering, or reputation-based services have limited success in addressing the constantly evolving nature of these attacks.
These methods often struggle to keep up with the surge of more tailored and customized phishing attacks, which occur in lower volumes and employ advanced social engineering tactics.
In response to this growing threat, recent advances in machine learning, particularly in  Natural Language Understanding (NLU), such as neural models based on the Transformer~\cite{vaswani2017attention} architecture, offer a promising avenue for combating such attacks.

In this work we describe CAPE, a modular system for BEC detection that has been deployed as part of an enterprise email security product and protects a large volume of mailboxes across a wide range of organizations.

Email is a complex data format encompassing multiple modalities such as text, images, attachments, semi-structured meta-data or the context of the communication (identities and relationships between involved entities).
Successful detection of BEC requires a system that combines information across all these modalities.
CAPE leverages a modular architecture that enables the combination of various approaches to create a robust system capable of evolving over time.
By focusing on modularity and the integration of machine learning approaches with domain knowledge for specific sub-problems, CAPE overcomes challenges posed by data sensitivity, labeled data scarcity, and practical limitations related to operational costs and the need to evolve the system over time.
    
A crucial feature of CAPE is its ability to provide interpretable verdicts by detecting a combination of higher-level signals.
This offers valuable insights for users, such as Security Operations Center (SOC) analysts, in their further investigations.
A key aspect of CAPE is its ability to adapt and improve over time by incorporating feedback and adjusting its detection mechanisms accordingly.
    
We provide an overview of CAPE's design principles, architecture, detectors, as well as a thorough description of processes related to efficacy evaluation, which are essential in practice.
We discuss various lessons learned from operating and improving CAPE for over two years and describe key factors that contributed to its ongoing success.

The paper is structured as follows.
In Sections~\ref{sec:background}~and~\ref{sec:related_work} we define what constitutes a BEC attack and summarise previously published related work.
In Section~\ref{sec:principles} we outline the key design principles that influenced the architecture described in Section~\ref{sec:architecture}.
Section~\ref{sec:detectors} provides an overview of a select sample of detectors in CAPE and Section~\ref{sec:classification} describes how detections, which constitute weak signals, are aggregated to convictions an how the system is being updated in a Bayesian setting.
Section~\ref{sec:efficacy} describes how the efficacy of the system and its parts is being tracked.

\section{Business Email Compromise}
\label{sec:background}
Business Email Compromise (BEC) can be broadly thought of as a social engineering attack targeting organizations via email~\cite{siddiqi2022study}. While BEC may vary in terms of sophistication, it is often hand-crafted and targeted~\cite{kotenko2022detection}, thereby more akin to spear-phishing rather than traditional phishing. Both phishing and BEC are subtypes of scam email~\cite{jakobsson2016understanding, almusib2021business}, or email fraud, albeit the scope of phishing surpasses email communication alone~\cite{thakur2014catching}. What differentiates BEC from phishing is that businesses or other institutions are the target entities for the attackers. Consequently, individuals influenced by the attacker serve primarily as the gateway to the organizational resources rather than their own. We define a successful BEC attack as one that leads a user affiliated with an organization to act upon a deceptive email, ultimately resulting in a loss of critical organizational resources, such as funds, data or access thereto, that would otherwise remain secure without the recipient's actions.

Classic interpretations of BEC perceive it as largely synonymous with email impersonation attacks in corporate contexts~\cite{mark2018becgang, zweighaft2017business}. According to the influential publication released by the US Federal Bureau of Investigation BEC may constitute one of the following~\cite{ic3InternetCrime, remorin2018tracking, mansfield2016imitation}: 
\begin{enumerate}[label=\arabic*)]
    \item an exploit of a known relationship between two organizations
    \item a CEO-fraud, i.e.~a request from a sham senior executive to make a wire transfer
    \item a payment request from compromised employee’s email account
    \item an attorney impersonation, i.e.~a fund transfer request from attackers, pretending to need to handle an urgent issue
    \item a data theft by business executive impersonation.
\end{enumerate} Notably, 1), 2), 4) and 5) pertain to impersonation whereas 3) refers to account compromise. Similar categories have been identified by other authors~\cite{jakobsson2016understanding, teeraknok2020practical}.
With the evolution of the attacks~\cite{dobrinoiu2021business}, more comprehensive definitions emerged, usually detaching the term from specific techniques. Literature offers diverse explanations of BEC's characteristics, however, there is a general consensus that it involves email communication, and most authors concur that it targets organizations and is typically motivated by financial gain or information acquisition, which coincides with the all-encompassing understanding of BEC as tantamount to ``email attacks on business systems''~\cite{buddhika2023detecting}.

The challenge of establishing the meaning for BEC is aggravated by the plethora of strategies implemented by malicious actors against businesses and other institutions~\cite{pierson2023preparing}. Such techniques vary greatly in terms of sophistication; from an attacker trying to build a relationship with the victim by exchanging several emails~\cite{cidon2019high}, through account compromise (i.e.~obtaining access to the email account of a victim)~\cite{ic3InternetCrime}, or tailored impersonations (of e.g., CEOs, other employees or affiliates, brands, vendors)~\cite{ic3InternetCrime, teeraknok2020practical} to less customized attempts that do not rely on impersonation, including email spoofing or masquerading~\cite{cidon2019high}. Regardless of their complexity, attacks often involve a request for a wire transfer~\cite{bakarich2020something}, a cryptocurrency payment~\cite{spangler2021business} or credential inputs. Such solicitation signals may permeate through various parts of the email. Most frequently, they are found directly in the email content, which may contain either authoritative and compelling, or reassuring and cordial natural language expressions, often in a tongue familiar to a victim, represented through text, images and images with text. Alternatively, an attacker may lure the recipient to follow a link directing to counterfeit or malicious website, or open and/or act upon the attachment (e.g., a form)~\cite{meyers2018not}. Recurrent elements among personalized BEC encompass credential theft and efforts to shift communication to other, usually less rigorously monitored, means, such as telephone calls~\cite{dash2023are} or private email accounts. Some techniques are not mutually exclusive; for instance, the impersonation of a vendor can be accompanied with a malicious attachment such as fake invoices.

\begin{figure}  
    \centering  
    \begin{tikzpicture}[  
    emailbox/.style={rectangle, draw, fill=white, text width=38em, rounded corners, minimum height=4em},  
    ]  
      
    \node [emailbox, font=\ttfamily] (content) {
        Sender: John Doe <jdoe@example.com>\\
        Recipient: Bob <bob@company.com>\\
        Subject: Help needed!\\~\\
        Good Morning,\\ let me know when you are unoccupied. I need you to get something done for me.\\ Thanks, John
    };  
      
    \end{tikzpicture}  
    \caption{An example of a BEC email that can't be convicted based on its content alone. To decide we need further information about the communication history between the involved entities and their identities.}  
    \label{fig:context_needed_example}  
\end{figure}

Due to the nature of more sophisticated and personalised techniques, it is often difficult to identify a BEC attempt based on an email alone.
For example, the sample BEC attack at Figure~\ref{fig:context_needed_example} is almost non-identifiable as BEC without contextual information such as whether there exists a previous communication between the sender and recipient, the recipient's identity, etc.
We call this set of information that is useful for classification but is not included in the email itself the email's {\it context}. 
We further elaborate on the context in Section~\ref{sec:context}.

% Social engineering
The key component in determining the effectiveness of a phishing or a BEC attack is whether the victim chooses to comply with fraudulent requests. The psychological aspects of social engineering in phishing and BEC attacks have been explored in previous studies~\cite{van2019cognitive, oliveira2017dissecting}. For instance,  Van der Heijden et al.~\cite{van2019cognitive} discuss and evaluate Cialdini's \emph{principles of influence}: \emph{Reciprocation, Liking, Social Proof, Authority, Scarcity, Commitment and Consistency}~\cite{cialdini1984influence} as a theoretical framework of persuasion for phishing attacks. \emph{Scarcity} (e.g.~insinuating that there is only limited time for some action such as resetting a password via a phishing link) has been found to be a major factor in complying with a fraudulent request~\cite{van2019cognitive}. Similarly, \emph{Authority} and its perceived symbols~\cite{cialdini1984influence}, such as brief, executive-like writing style, explain the prevalence of impersonation~\cite{distler2023influence}, particularly of CEOs, in BEC attack methods. On the contrary, \emph{Reciprocation} (i.e.~the sense of obligation to return a favour) is considered to be less useful in business environments, likely due to a greater awareness of potential conflicts of interest among recipients~\cite{van2019cognitive}. Other principles, such as \emph{Commitment and Consistency}, while not as influential, might be applicable in more long-term strategies, e.g.~establishing rapport or sending follow-up requests, which have a higher likelihood of success given the victim's established relationship with the sender. These principles serve as tactics for exerting pressure on individuals. The effect is exacerbated in a professional setting, as employees are naturally inclined to respond to inquiries they believe are work-related~\cite{distler2023influence}.
Despite possessing theoretical knowledge gained from anti-phishing trainings~\cite{williams2018exploring}, victims (in the fashion of Cialdini's \emph{automatons}) may still unintentionally overlook their own vulnerability~\cite{cialdini1984influence}. Hence, blocking unwanted messages before they reach a user is a critical security measure.

\section{Related Work}
\label{sec:related_work}

The growing complexity and continuous evolution of BEC and spear phishing demand adaptable machine learning systems that can respond to new data, including emerging attack types. Typically, features (or \emph{signals}) are either engineered from the email content, for example, by key-word matching~\cite{ho2019detecting}, detecting suspicious links via domain popularity or URL analysis~\cite{cidon2019high}, traditional statistical natural language processing methods such as TF-IDF~\cite{harikrishnan2018machine, cidon2019high} or basic topic-modelling methods such as Latent Dirichlet Allocation~\cite{shyni2016multi}. 
Besides features engineered from the email's content, metadata can be used (e.g., sender, receiver, CC fields), for instance to determine domain popularity or detect impersonation~\cite{cidon2019high}. 

Many classical machine learning approaches, that were also commonly used for spam classification in the past \cite{guzella2009review}, have been employed for BEC detection using the aforementioned features: random forests for BEC~\cite{cidon2019high} or phishing detection~\cite{ho2019detecting}, $k$-nearest neighbors~\cite{cidon2019high}, support vector machines (SVM)~\cite{shyni2016multi, harikrishnan2018machine} and logistic regression~\cite{harikrishnan2018machine}. In~\cite{cidon2019high}, the architecture of \emph{BEC-Guard}, another productionized engine, is outlined. This engine employs, among other content and metadata-based classifiers, a link classifier using three features, namely domain registration date, URL field length, and domain popularity, which are used to train a random forest. 

Different deep learning and large language models (LLMs) models can be found in recent work~\cite{halgavs2020catching, lee2020catbert, benkovich2020deepquarantine}. In~\cite{halgavs2020catching}, for instance, an LSTM-based model is used to detect phishing. \textsc{Catbert}, a further end-to-end solution, uses \textsc{Bert}~\cite{devlin2018bert}, a transformer architecture, to detect social engineering from emails using both the email's content and metadata. These methods are end-to-end and do not rely on feature-engineering. \textsc{Lime}~\cite{ribeiro2016should}, a model-agnostic technique to provide a certain level of explainability, can be used together with end-to-end models. For instance, the contribution of input tokens for a prediction task (e.g.~BEC vs.~benign) can be examined~\cite{lee2020catbert}. While this may facilitate a certain level of interpretability, the internals of such models remain a black box. The lack of such \emph{model transparency} together with high inference costs is a limitation that needs to be considered in production systems~\cite{lee2020catbert}. 

Certain studies report performance metrics on the basis of non-publicly accessible datasets~\cite{duman2016emailprofiler}, at times from deployed models~\cite{lee2020catbert, cidon2019high}, while others adopt emails from public corpuses exclusively~\cite{halgavs2020catching, harikrishnan2018machine}. Since public corpora are not updated regularly, they are not representative of the prevailing BEC landscape. Dataset sizes used for model training and validation differ in orders of magnitude: from ten thousands~\cite{duman2016emailprofiler} to hundred of thousands~\cite{stringhini2015ain} to millions of emails~\cite{lee2020catbert}.

\section{Design Principles and Practical Constraints}
\label{sec:principles}

In this section, we outline the constraints and goals that influenced the design of our detection architecture.
These range from the nature of data we process, operational and cost considerations to the need to maintain and evolve the system over time.

CAPE is part of an enterprise product that includes other engines covering various aspects of email security such as phishing and spam protection, analysis of attachments, URLs, etc.
This allows~CAPE to focus on detection of BEC and sophisticated phishing attacks.
The guiding principle is that CAPE does not include specific rules targeting concrete malicious emails but focuses on detecting behaviors that are shared and generalize to future attacks.

The major constraint we need to work with is \emph{data sensitivity}.
Emails are highly confidential and thus research access is limited by strict privacy policies.
This limits both the size of data that can be stored and its retention period.
Research access is generally possible only to a small subset of emails that are highly suspicious or to emails reported as false negatives.

Furthermore, BEC and sophisticated phishing emails are extremely rare which makes the classification problem severely class-imbalanced.
The true prevalence of BEC emails is difficult to quantify, as many incidents go unreported but our estimate is in the range from $10^{-4}$ to $10^{-5}$ depending on concrete deployment environment.
The estimate is in line with previously published work~\cite{cidon2019high}.
This means the system needs to have extremely low false-positive-rate $(\mathrm{FPR} = \frac{\mathrm{FP}}{\mathrm{FP}+\mathrm{TN}})$ to provide convictions with high precision  ~$(\frac{\mathrm{TP}}{\mathrm{TP}+\mathrm{FP}})$~\cite{brabec2020model}.
Not having research access to regular benign emails, which form the vast majority of the traffic, makes this task not suited for a fully end-to-end ML solution.
This is further complicated by unavailability of representative labeled data even for the positive (BEC) class.
The publicly available email datasets are either old, focused on less severe threats~\cite{spamassassin_corpus,trec2007_spam_public_corpus,nazario2007phishing} or not related to security~\cite{klimt2004introducing}.
While potentially useful for testing and development of proofs of concepts, the utility of these datasets is very limited compared to production data.
Other datasets and feeds consist of indicators of compromise (IOCs) such as URLs or attachment file hashes~\cite{phishtank,openphish}.
Such feeds play an important role in email security products and when applied carefully can be used to label data to an extent.
The problem is that, in our experience, they suffer from a relatively high amount of false positives and don't help much in BEC detection as such attacks often do not contain these IOC types.

Thus, in the beginning we were faced with a \emph{cold start problem} having insufficient data and labels to train an ML model that would operate at higher levels of abstraction, for example classify emails based on their full contents.
Any such attempt would inevitably lead to learning shortcuts~\cite{geirhos2020shortcut} that do not generalize to production data.
Instead, we designed the system in a way that allows combination of domain knowledge and ML.
We limited application of ML to situations where we possess representative data, for example development of a solid \emph{urgency detector} is possible with data fully obtained from public sources such as Enron dataset.

To resolve the cold start problem it is essential to iterate quickly and continuously improve the system based on its performance on real production data.
The system can run silently at first until efficacy is sufficient.
The main efficacy measure of the system is precision which needs to be extremely high and maintained at all times.
Precision acts as a constraint.
Recall~$(\frac{\mathrm{TP}}{\mathrm{TP}+\mathrm{FN}})$, on the other hand, is a measure that is being increased over time.
In practice, it is impossible to measure recall because the number of false-negatives is unknown.
As a proxy, if we consider precision to be fixed, we can measure and aim to increase conviction-rate~$(\frac{\mathrm{TP}+\mathrm{FP}}{\mathrm{TP}+\mathrm{FP}+\mathrm{TN}+\mathrm{FN}})$ over time.
To continuously evolve the system, we need a modular architecture that is easily extensible and an ability to measure performance of individual parts.
Over time, we gathered more representative data enabling broader applications of ML than in the beginning.

Emails convicted by CAPE are in some cases being investigated by a SOC team or us for verification or to identify larger attack patterns.
For this use-case, it is extremely desirable that the verdicts are \emph{explainable}, allowing the team to understand why the system made a conviction.
As we describe in greater detail in Section~\ref{sec:architecture}, the modular architecture provides solid explainability of the verdicts.

Another constraint that is taken into account and influences the architecture are \emph{operating~costs} and most importantly variable costs that scale with the amount of emails processed.
CAPE receives raw emails on input which are on average 300~KB in size including headers.
Applying large NLU models such as LLMs indiscriminately on every email we process would be beyond our cost envelope.
To simplify, the costs are primarily affected by the required throughput of the system and the type and amount of hardware (in our case cloud compute instances, databases, etc.) it needs to provide such throughput.
Latency is a constraint and there is always an upper limit. 
For CAPE, the maximum allowed latency to handle a single email is in the range of multiple seconds which is reasonably flexible.
However, using this computational budget for every email would either decrease throughput or increase hardware needs leading to high operating costs.

We use the fact that vast majority of emails are benign and for most of them the decision is easy without costly analysis.
Thus the detection pipeline is designed to processes easy emails fast.
This leaves more computational budget to spend on a small number of interesting emails.
This is reflected by CAPE's email processing time distribution with 50 \% of emails being processed in 50 ms, 75 \% below 100 ms and 99 \% below one second.

\subsection{Email Context}
\label{sec:context}

\begin{figure}[h]
\centering
\includegraphics[width=0.6\textwidth]{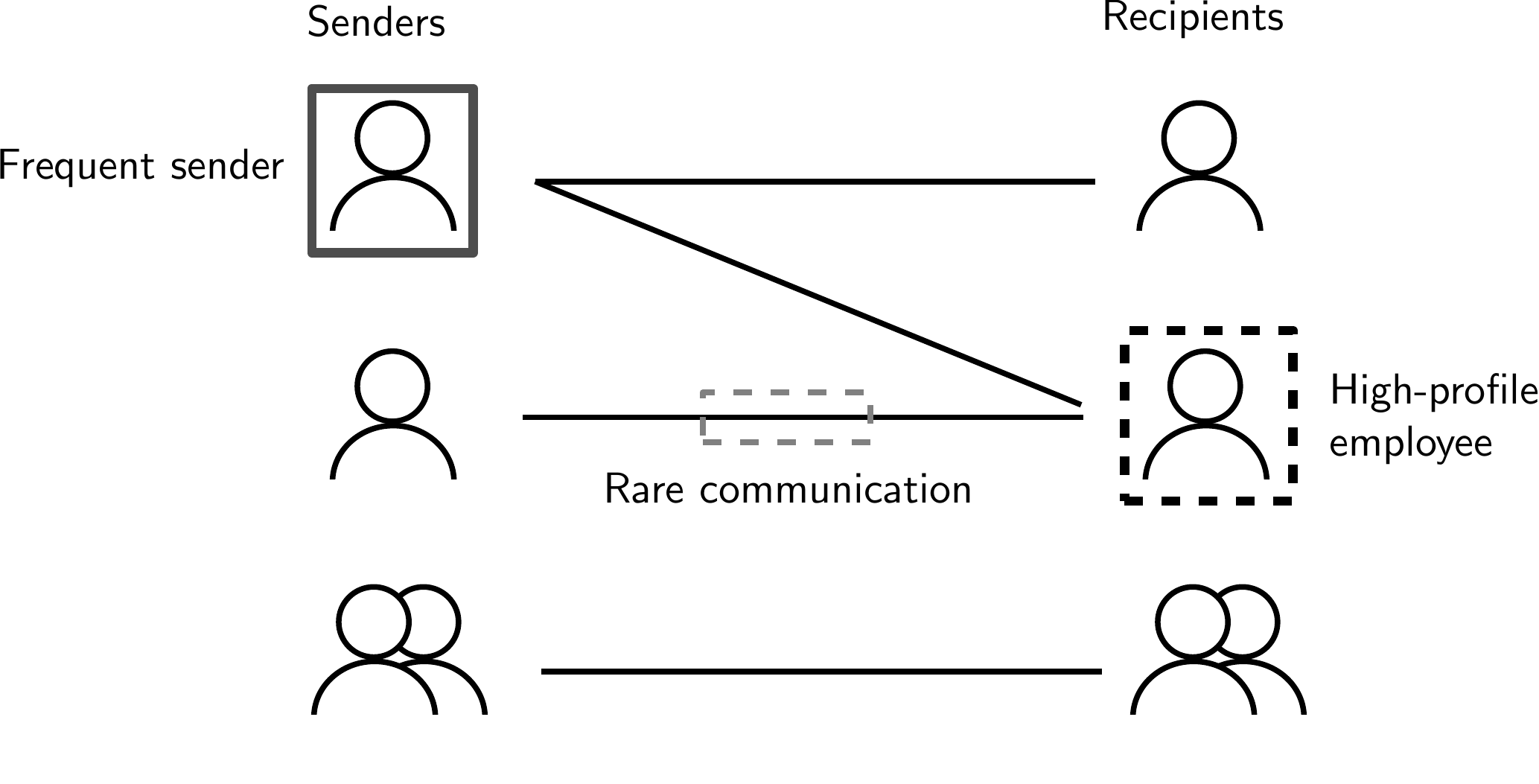}
\caption{\emph{Mail Graph} provides contextual information about identities and enables the identification of common and rare communications between entities. Detectors based on this information are described in Section~\ref{sec:mg_detectors}} \label{fig:mail_graph}
\end{figure}

Figure~\ref{fig:context_needed_example} depicts an example BEC email that can't be convicted based on its content alone and requires a broader context.
We provide contextual information to our classification pipeline via a component titled \emph{Mail Graph}. 
In Mail Graph, depicted in Figure~\ref{fig:mail_graph}, we model identities and relationships of sending and receiving entities.

The component provides single-digit millisecond latency read access to historical patterns. 
The identity and relationship models are continuously updated asynchronously and efficiently by batching and aggregating to reflect changes in behaviors.
Mail Graph is modelling the data on three different levels - global, customer/business specific and user specific.

\section{Detection Architecture}
\label{sec:architecture}

\begin{figure}[h]
\centering
\includegraphics[width=0.8\textwidth]{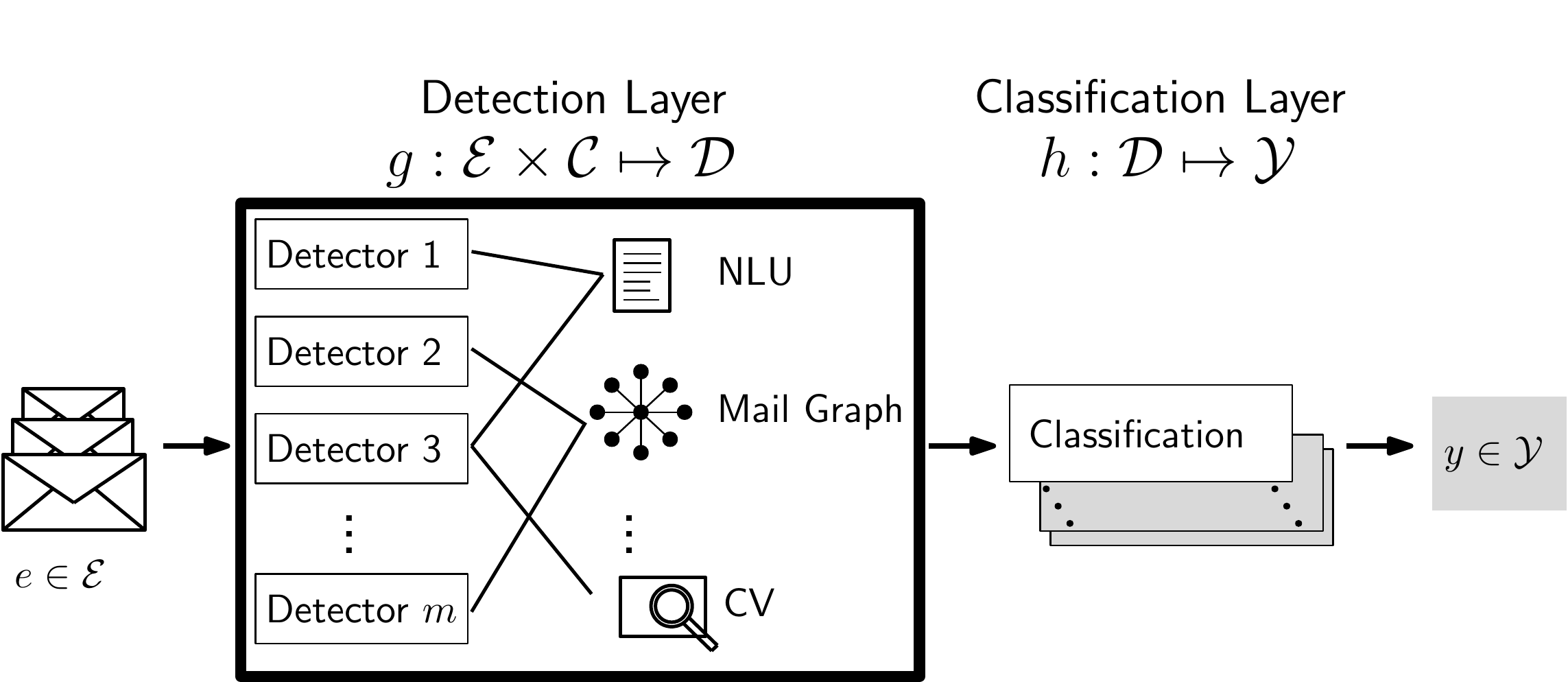}
\caption{{\it Classification pipeline} consists of two layers: 1) detection layer and 2) classification layer. Emails are processed independently to allow horizontal scaling of the system. The detection layer features a heterogeneous set of detectors, targeting various email modalities, aiming to identify behaviors correlated with BEC attacks. The classification layer classifies the email based on the identified detections.} \label{fig:cape_architecture}
\end{figure}

Given the requirements and constraints, we propose two-layer architecture depicted at Figure~\ref{fig:cape_architecture}.
The \emph{detection layer} is modular, comprising of independent detectors focused on detecting behaviours and attack techniques in the email itself and its communication context.
The detections are often weak signals by themselves.
For example, \emph{urgency} which is one of the detected techniques occurs frequently in benign emails as well as in malicious.
The detections are combined inside a \emph{classification layer} which provides the final verdict and its confidence.
This decomposition into independent detectors makes the verdicts explainable as instead of just a binary verdict, we are able to provide the set of behaviors that makes the email suspicious and highlight them in the user interface. 
The encapsulation and separation of concerns allows for parallel research and development while easing maintenance in the long term.
Both layers are described in greater detail in the following sections.

The two layers form a \emph{classification pipeline}.
Formally, the classification pipeline is a function $f: \mathcal{E} \times \mathcal{C} \to \mathcal{Y}$, where $e \in \mathcal{E}$ is a single email on input, $c \in \mathcal{C}$ is the \emph{email's context} from Mail Graph and $y \in \mathcal{Y}$ is the final verdict.

In practice, the final verdict $y \in \mathcal{Y}$ is accompanied by metadata consisting of the detections that are created in the detection layer.
The formal definition of the pipeline above is a simplified version that omits the metadata passing as it is technically trivial and obscures the notation.
The metadata are propagated all the way to the user interface and are essential for the aforementioned explainability reasons as they provide insights into why the email was convicted.

The detection layer consists of a set of independent detectors that may or may not yield a detection.
Each detection consists of a score and metadata.
The precise semantics of the score $d_s \in [0, 1]$ depend on the specific detector but in general the scores represent the detector's confidence in the detection and sometimes also the detection's severity (e.g. a score can be lowered if the detection is likely to be in a marketing context such as an \emph{open redirect}~\cite{cwe601} detection leading to a known marketing service).
The metadata encompasses any information from the email or its context that are valuable to provide with the detection.
For example, an \emph{urgency} detection contains a segment of the email message where urgency was detected and the open redirect detection includes the inner and outer URLs in its metadata.
The structure of metadata varies by detection type.

Several detectors serve a special purpose and target signals that strongly indicate the email as benign.
If such detector yields a detection it stops the execution of the remaining detectors in the detection layer which saves computational resources.
The remaining detectors are assigned zero score and skipped.

We omit the metadata from the notation and describe the detection layer as a function $g: \mathcal{E} \times \mathcal{C} \to \mathcal{D}$ having the email and its context as input and a vector of detection scores $\mathcal{D} = [0,1]^m$ as output, where $m$ is the number of existing detection types.
The $i$-th dimension of $\mathbf{d} \in \mathcal{D}$ is equal to the score of the $i$-th detection type or zero if detection of this type was not yielded by any detector. The classification layer is a function $h: \mathcal{D} \to \mathcal{Y}$ and thus the entire pipeline can be expressed as $f = h \circ g$.

\section{Detectors}
\label{sec:detectors}

A detector is an independent unit that recognizes specific behaviors and patterns.
The complexity of detectors ranges from classic algorithmic approaches to large neural network models. 

Every detector has access to a parsed email and to email context.
The raw email is parsed according to established standards, but exceptions occur as incoming emails do not always adhere to these standards.
In real traffic, nearly any kind of malformed header or HTML can be observed.
In such cases, we follow a rule of thumb and treat the email similarly to how an email client used by a user to view the email would handle it.
Detectors can also analyze images embedded in the emails and text from images is extracted via OCR.

Due to the nature of the data, 
The detectors target both strong and weak signals.
A weak signal is a detection not indicative of BEC by itself but only in combination with other detections.
Thus many individual detections occur in benign traffic, and the primary focus of detectors is not to solve the ultimate task of detecting malicious emails.
The goal is to detect behaviors likely to generalize to a variety of attacks, including those not yet observed, as some highly targeted attacks may only be observed once.
This approach contrasts with finding concrete indicators of compromise targeting an ongoing campaign.

A small portion of the detectors is targeting benign signals in the email or its context.
If such signals are identified with high confidence, the detection layer may decide to stop further analysis.
This reduces the average processing time and leaves more resources for analysis of more suspicious emails.

CAPE currently includes over 90 detectors.
For documentation purposes, we track all of these using \emph{model cards}~\cite{Mitchell_2019}, which summarize information such as detector name, description, status, metadata, supporting infrastructure, measured efficacy, training data and potential caveats. 
Below, we provide a high level summary of a few selected detectors.

\subsection{Language Model-based Detectors} 

We employ several content detectors which analyze segments of the email body.
A segment can either be a sentence or a paragraph.
{\it Call-to-action} detector identifies natural language request and recognizes prompts to perform various potentially dangerous actions, such as opening an attachment, clicking a link or disclosing sensitive data.
We are also detecting requests for recipients credentials, or urgency in the email.

These detectors are constructed using pre-trained Transformer-based models~(LLMs)~\cite{vaswani2017attention}, which are subsequently fine-tuned in a supervised manner on a manually labeled dataset.
To improve computational efficiency, these models employ a multi-label approach, enabling them to simultaneously output multiple suspicious signals.

Because of significant resource requirements for inference, we need to reduce the number of segments that are analyzed with such a large model.
For this, we utilize a layered filtering mechanism.
First, we use a small, weaker model trained to quickly identify sentences likely containing call-to-action.
This model is tuned to have high recall at the cost of fairly low precision, and serves as a pre-filter for the accurate, but demanding Transformer-based model.
The pre-filter model currently filters out approximately 75\% of segments.
The training of the pre-filter model is done using a student-teacher approach utilizing labels predicted by the larger model.
As such training of the pre-filter model can be automated and does not require manual labeling.
The pre-filtered segments are classified by the Transformer-based model.

Starosta~\cite{radek2021detekce} discusses the Call-to-action detector and methods to efficiently deploy it at scale in detail.
Due to ongoing progress in Transformer architectures, the underlying models are being regularly updated.
Call-to-action and urgency are weak signals by themselves, commonly occurring in benign emails.

\subsubsection{Automated Labeling Using Generative LLMs}

Development of the above-mentioned content detectors requires creation of a high-quality dataset of labeled sentences or longer text segments.
Obtaining such dataset requires labor-intensive manual labeling.
Due to data sensitivity (even though publicly available data can be helpful to an extent) and the fact that the labeling task generally requires solid knowledge of the domain, outsourcing the labeling is not a viable option.

To ease development of similar detectors, we have since started utilizing the GPT-4~\cite{openai2023gpt4} model to automatically label training data.
Querying generative models in production to analyze millions of emails per day would not be feasible due to both cost and latency.
However, we can utilize them offline to automatically classify data from a large dataset of emails, which can then be used to train a more cost-efficient classifier.
Although current generative models do not provide human-level classification accuracy, given a large enough dataset we are able to produce a capable model from the automatically labeled data in a fraction of time.

\subsection{Email Address Masquerade Detector}
\label{sec:address_masquerade}
This detector aims to detect, when the sender is masquerading his real email address by a different one, which is more trustworthy for the recipient. This is achieved by displaying a fake email address in the sender display name. Note that there are almost no restrictions on the sender name format in the email infrastructure.
See a typical example below:
\begin{center}  
\texttt{\char`\"John Doe john.doe@example.com\char`\"~\textless noreply@evil.name\textgreater}  
\end{center}  

There exists a variety of MIME-encoding tricks the attacker can use to disguise the above and that we detect.

This detection type is an example of a seemingly strong signal, which is also found in real legitimate traffic in a considerable quantity of emails from legitimate services that "impersonate" the user for benign reasons.

\subsection{Unicode Masquerade Detector}
In order to evade detection, some attacks use homoglyphs~\cite{holgers2006cutting}, i.e. lookalike symbols from different scripts, which can be elusive for various natural language processing techniques.
However, usage of such masqueraded symbols and mixture of different scripts can be considered a signal of maliciousness on its own.
For example, consider the sentence below:

\begin{center}
You h$\cya$ve a new inv$\cyo$i$\cys$e. Ple$\cya$se  make they $\cyp\cya$yment by end of d$\cya$y.
\end{center}

It looks normal to a human eye, but it is not meaningful, when looking at the code points, because Latin script was interleaved with lookalike Cyrillic (namely \textcyrillic{a}, \textcyrillic{c}, \textcyrillic{o}, \textcyrillic{P}) that have a different meaning after transliteration (e.g. \textcyrillic{P} $\mapsto$ R).

Another common technique is abuse of zero-width spaces, where the attacker distorts the words with invisible characters such as the Unicode soft hyphen (U+00AD), making classic engines not recognize the word "password" below:

\begin{center}  
\texttt{p\textcolor{gray}{\textbackslash u00ad}\textcolor{gray}{\textbackslash u00ad}as\textcolor{gray}{\textbackslash u00ad}\textcolor{gray}{\textbackslash u00ad}s\textcolor{gray}{\textbackslash u00ad}\textcolor{gray}{\textbackslash u00ad}wo\textcolor{gray}{\textbackslash u00ad}\textcolor{gray}{\textbackslash u00ad}r\textcolor{gray}{\textbackslash u00ad}\textcolor{gray}{\textbackslash u00ad}d}  
\end{center}

All these techniques provide a potentially useful signal to detect. However, there are still valid use cases, such as beautified text or languages mixing scripts in normal text.

\subsection{Communication Frequency Detectors}
\label{sec:mg_detectors}

Detector of {\it rare communication} between sender and recipient utilises the email context from Mail Graph.
The detector considers various potential relations between the sender and recipient such as whether the sender address is rare for the whole recipient company, or possibly only for an individual recipient.

The detection logic considers both the quantity of the emails exchanged in the past and their distribution across a historic window to determine the exact detection score.
A connection that occurs repeatedly for multiple days is considered more stable than a single-day burst of emails.

The opposite of the rare communication detector is the {\it frequent communication} detector which works on a similar principle.
Past frequent communication between entities is one of the strong signals that email is likely benign.
The detector learns the commonly used user-specific and customer-specific services automatically and eliminates the need to maintain them in a static form.
\section{Classification}
\label{sec:classification}

The classification layer ($h: \mathcal{D} \to \mathcal{Y}$) takes a detection score vector as input and yields a binary outcome.
We adopt $\mathcal{Y} = \{0, 1\}$, with $0$ denoting benign and $1$ representing the BEC (positive) class.
This is a conventional learning setting and if a sizable, representative, labeled dataset was available then a variety of supervised learning methods could be used to obtain a model.
Such dataset, however, does not exist.
The data-related limitations and the cold start problem were described in Section~\ref{sec:principles}.
These limitations determine what types of models and learning algorithms can be used.

In short, the labels are scarce and affected by a selection bias, with particularly the benign class not being sampled well, and the classification problem is severely class-imbalanced.
Learning any model in this setting from the data alone is deemed to generalize badly by learning shortcuts~\cite{geirhos2020shortcut} and overall resulting in excessive false positive rates when tested in production traffic.
To solve this problem, we need to combine \emph{domain knowledge} with the data, as the information necessary for successful generalisation is not available in the data alone.
Domain knowledge, encompasses information we have about the threat landscape, deeper understanding of performance of individual detectors from their targeted evaluations and also knowledge of various fixes and adjustments to the detectors that would reflect in the data only after a delay.
We also do not posses any past data for newly developed detectors.
All this makes domain knowledge essential for practical outcomes.

In addition, we are not concerned with finding an optimal model as that is practically intractable and the data evolves over time.
Instead, we approach the task iteratively, aiming to continuously improve the classification layer over time.
In this setting, a new model is not required to be optimal but it suffices to be better than the currently deployed one.

The need to combine data with domain knowledge means we require the classifier to be truly interpretable.
While defining what makes a model interpretable is complex~\cite{lipton2018mythos}, it generally holds that smaller models that perform less steps during inference are preferred.
We now compare two potential model types: 1) tree-based models; 2) generalized linear models.

Tree-based models such as decision trees could be used successfully to perform the classification layer's task.
Decision trees are expressive, interpretable if reasonably small and their structure can be learned from data and successively manually adjusted to inject domain knowledge.
We do not use decision trees, because they are harder to gradually evolve.
Their structure is a non-smooth function which is less suited for the Bayesian learning paradigm we describe later.
Also, adjusting a tree to include a new type of detection might require modification at many sub-trees or rebuilding the tree's structure from scratch.

We currently use a single \emph{logistic regression} model to fulfill the classification layer's function.
The model is defined as:

\begin{equation}
\label{eq:logreg}
h(\mathbf{d};\mathbf{w},t) = [\mathrm{logistic}(\mathbf{w}^\top\mathbf{d}) \ge t] = \left[\frac{1}{1 + e^{\mathbf{w}^\top\mathbf{d}}} \ge t\right]
\end{equation}

Despite the linear decision boundary and smaller capacity compared to decision trees, the model exhibits a solid level of expressivity due to the higher-dimensionality of the data~\cite{gorban2018blessing}.
The model and its decisions are simple to interpret, because each detection type is associated with a just a single scalar weight.
The higher the weight the more influential the detection is.
Thanks to the smoothness of the logistic function, two models $h_a$ and $h_b$ are compared easily by analysing weight differences $\mathbf{w_a} - \mathbf{w_b}$, which is an extremely practical property when updating the model. When a detector is updated or added, it is often sufficient to adjust its weight in isolation.

In the remainder of this section we describe how the parameters of the logistic regression model are learned.
Before that, we would like to highlight that the classification layer is designed to potentially support multiple independent classifiers.
Given $k$ classifiers, the layer's output would be given by $h(\mathbf{d}) = \max_{i=1}^k h_i(\mathbf{d})$.
A classifier can either be general, as is currently the case or focused only on certain sub-problems such as credential phishing or financial frauds.

\subsection{Bayesian updating of the logistic regression model}
\label{sec:bayesian}

As the data evolves and detectors are being added or modified, the model is being continuously updated in an iterative fashion.
Due to the absence of a labeled dataset representative of current production data for validation before deployment, we rely on the successive models' weights being reasonably close to each other, which prevents drastic unforeseen performance shifts between models.
This approach resembles the use of a conservative learning rate during gradient descent.

In the beginning, when faced with the cold start problem, the initial model was learned on a combination of public datasets mentioned in Section~\ref{sec:principles}.
This provided the initial estimate of the parameters, but multiple iterations of the model on actual production data and successive manual adjustments of the parameters were required to arrive at a model with acceptable precision and recall.
These manual adjustments can be seen as a way to inject domain knowledge, not present in the training data, into the model.
At this stage, with no practical model existing yet, it was necessary to adjust the parameters in the context of each other.
On the other hand, compared to the present day, a much smaller number of detection types existed, which simplified the problem.

Apart from adjusting the weights manually, which is a fast reaction to larger changes in detector implementations or data, we employ a \emph{Bayesian framework} to evolve the model through a combination of domain knowledge and limited feedback data.

In this approach, the input is an already existing baseline model with weights $\mathbf{w_b}$ and a dataset $\mathbf{D}~=~\{(\mathbf{d_i}, y_i)~|~ \mathbf{d_i} \in \mathcal{D}, \, y_i \in \mathcal{Y}, \, 1 \leq i \leq N\}$.
The dataset is limited in size and suffers from various selection biases.
Particularly the negative class is not represented well.
The goal is to find a model with weights $\mathbf{\hat{w}} \in \mathbb{R}^m$, where $\mathbf{\hat{w}} = \arg\max_{\mathbf{w}} p(\mathbf{w}|\mathbf{D})$.
Using Bayes' theorem we get posterior distribution $p(\mathbf{w}|\mathbf{D}) \propto p(\mathbf{D} | \mathbf{w}) \cdot p(\mathbf{w})$ with $p(\mathbf{D} | \mathbf{w})$ and $p(\mathbf{w})$ being the likelihood and prior respectively.
The prior distribution allows us to inject domain knowledge into the estimation process.
We define a Bayesian logistic regression model as follows:

\begin{alignat*}{3} 
\mathbf{w} &\sim \mathcal{N}(\mathbf{w_b}, \mathbf{\Sigma}) & \\
p_i &= \mathrm{logistic}(\mathbf{w}^\top \mathbf{d_i}), \quad  &1 \leq i \leq N \\
y_i &\sim \mathrm{Bernoulli}(p_i),                    \quad &1 \leq i \leq N 
\end{alignat*}

The baseline model's weights $\mathbf{w_b}$ are used together with a covariance matrix $\mathbf{\Sigma}$ to define a normal prior distribution.
Matrix $\mathbf{\Sigma}$ is a hyperparameter and for simplicity we restrict ourselves to diagonal matrices.
Bernoulli distribution is used for the likelihood, which is the natural choice in Bayesian logistic regression.
We have the option to either sample from the posterior distribution $p(\mathbf{w}|\mathbf{D})$ using MCMC methods or, if only interested in a point estimate, obtain $\mathbf{\hat{w}}$ through gradient descent.
The Turing.jl package~\cite{ge2018turing} is utilized to define the model and determine the estimate $\mathbf{\hat{w}}$.

\begin{figure}[h]
\centering
\includegraphics[width=0.8\textwidth]{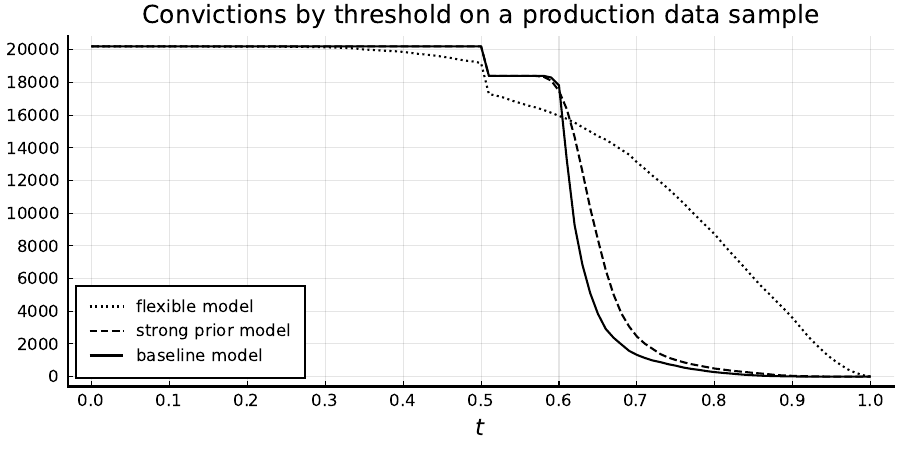}
\caption{Comparison of models learned via the Bayesian approach on a random dataset of 20000 weakly-suspicious emails sampled from the production traffic. The emails are unlabeled. The threshold from Equation~\eqref{eq:logreg} is on the x-axis. Compared to \emph{baseline model}, the \emph{flexible model} improved the area under ROC curve (AUC) by 0.153 on the test sibling of dataset $\mathbf{D}$. The \emph{strong prior model} improved AUC only by 0.035. However, due to biases in $\mathbf{D}$, the flexible model is overconfident and would convicts too many emails in production to be practical. The strong prior model, on the other hand, provides a modest improvement on $\mathbf{D}$ but behaves reasonably on production data.} \label{fig:convictions}
\end{figure}

Small values in $\mathbf{\Sigma}$ make the prior distribution stronger, forcing the posterior closer to $\mathbf{w_b}$.
Larger values in $\mathbf{\Sigma}$ give the model more flexibility to fit the dataset.
There is a trade-off between closeness to the baseline model and predictive performance on $\mathbf{D}$.
This trade-off is demonstrated in Figure~\ref{fig:convictions}.
Practical values of $\mathbf{\Sigma}$ are selected empirically and can incorporate knowledge about the detectors such as priors for established detectors being stronger.

\section{Efficacy Evaluation and Monitoring}
\label{sec:efficacy}

%There are two primary goals of Efficacy Evaluation and Monitoring. The first is understanding of the engines' efficacy in order to be able to balance precision and recall and also to understand the landscape changes over the time. And secondly, there is a goal of gathering samples of FPs and FNs from the traffic in order to guide the design and engineering process of enhancements and roadmap in directions needed for our customers. It is an essential element of the continous learning process that is required for the engines ability to reflect changing types of threats and their prevalence in the traffic.
%Both of the goals have a built in element of a feedback loop that by principle needs to be continous so that it keeps the ability to bring new insights over the time and prevent natural tendency of result deterioration if the systems were left without ongoing updates.

The dynamic nature of the environment necessitates ongoing efficacy monitoring in production to sustain and improve predictive performance of the system.
There are two primary factors that could potentially lead to a decline in efficacy.
Firstly, the data distribution is drifting over time.
The drift can be gradual as the style and typical content of emails changes over longer periods of time, or it can be sudden. For instance, sudden changes in data distribution may appear after onboarding new customers.
The second factor is the introduction of a new version of the system. Although this is a controlled change compared to data drift, there can still be a considerable unexpected impact on efficacy.

The most important metrics monitored on an ongoing basis are precision, conviction-rate (i.e proportion of positive verdicts) and the rates of false positives and false negatives received as customer feedback.
We described the rationale for these metrics and their treatment in Section~\ref{sec:principles}.
In addition, we track various data properties, such as the distributions of email content length, number of sentences, number of URLs, number of and types of payloads, languages distribution and more. 
For detectors specifically, we measure their impact on the system as described later in this section. 
All the mentioned metrics are monitored automatically using alerts when anomalies occur and periodically via dashboards. 

In the remainder of this section we elaborate on several topics related to efficacy monitoring.

\subsection{Customer feedback}

After overcoming the cold start problem,  we now have access to a feed of emails submitted as false positives or false negatives by our customer base.
The feed has limitations, such as a certain level of noise and non-uniform sampling as the proportion of provided feedback differs by customer. 
Nonetheless, such source of labeled data is precious.
In general, false positives are prioritized over false negatives as they are a result of an active action of the system and maintaining high precision is essential.
False negatives are a source of ideas for improvements such as new detectors and are also one of the sources (together with false positives) useful for Bayesian updates of the classification layer described in Section~\ref{sec:bayesian}.

\subsection{Rotating team efficacy evaluation duty}

To get an unbiased estimate of precision, it is necessary to label an i.i.d.~sample of convicted emails.
Due to the i.i.d.~requirement, it is not possible to use the customer feedback for this purpose.
We introduced an internal team process, which assigns one team member to analyze random 10-50 positively labeled emails each business day.
Such quantity is enough to estimate precision reasonably well.
The main purpose of this process is to label the emails to enable measurements and gather data for Bayesian updating of the classification layer.
A noteworthy side effect of this process is that it significantly deepens our understanding of the data, provides an essential feedback loop and ideas for improvements. 
Therefore, even though the overall efficacy operations include multiple teams, it is essential for the development team to remain involved and maintain regular contact with the system's output.
Appropriate internal tooling, such as email clustering to effectively label larger campaigns, has been developed to ease the evaluation.

\begin{figure}[h]
\centering
\includegraphics[width=0.6\textwidth]{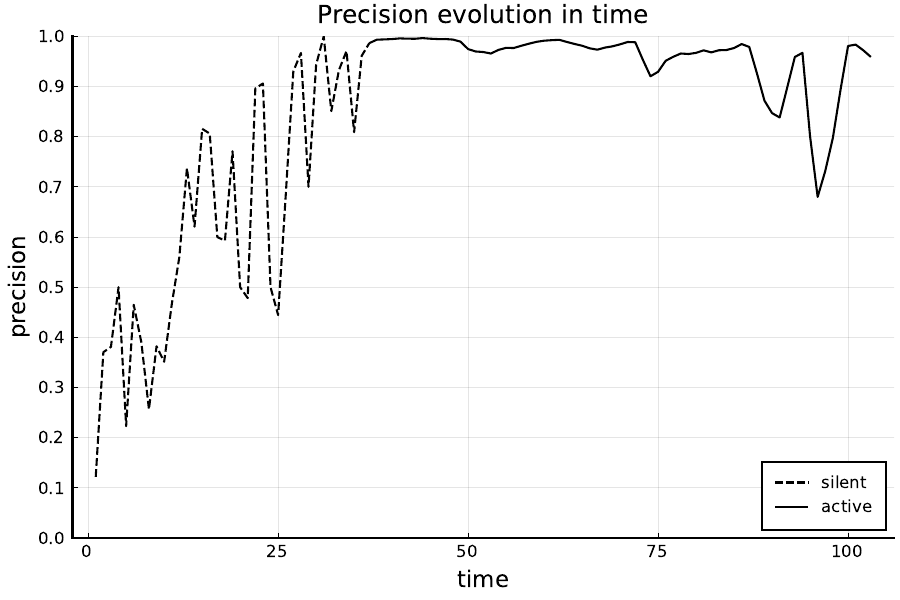}
\caption{Evolution of CAPE's precision over time with the measurements obtained through the rotating efficacy evaluation process. On the x-axis, measurements are chronologically ordered, with non-uniform distances between successive values, averaging one week apart.
Thus the measurement interval spans over two years.
Initially, CAPE was deployed on production traffic in silent mode, where its convictions had no effect. This period was essential for stabilizing the precision above 80\%, a level that has been consistently maintained over an extended period. The single drop below 80\% precision, has been discovered and swiftly remediated thanks to the regular evaluation process.} \label{fig:precision_plot}
\end{figure}

The Figure~\ref{fig:precision_plot} depicts evolution of the precision in time with measurements obtained through this process.

\subsection{Quantifying detector impact}
To measure the performance of individual detectors in an automated unsupervised way, we use the following two metrics.
First, we define \emph{relative impact}, which is calculated by counting the relative number of positively labeled emails that would not be convicted if the detector didn't exist.
Detector $i$ is said to be impactful for a given email, if the classification layer $h: \mathcal{D} \mapsto \mathcal{Y}$ produces a positive verdict using all detectors $h(\mathbf{d})=1$, but would produce a negative verdict if detector $i$ was removed $h(\mathbf{\bar{d}})=0$, where 
$$
\mathbf{\bar{d}} = (\bar{d}_1, \bar{d}_2, \dots, \bar{d}_m)
$$
$$
\bar{d}_j = 
\begin{cases}
    0, & \text{if } j = i \\
    d_j, & \text{if } j \neq i
\end{cases},
$$

hence for a single email:

\begin{equation}
\label{eq:impact}
\mathrm{impactful}(i; \mathbf{d}) = [h(\mathbf{d}) = 1 \wedge h(\mathbf{\bar{d}}) = 0]
\end{equation}

\begin{figure}[h]
\centering
\includegraphics[width=0.6\textwidth]{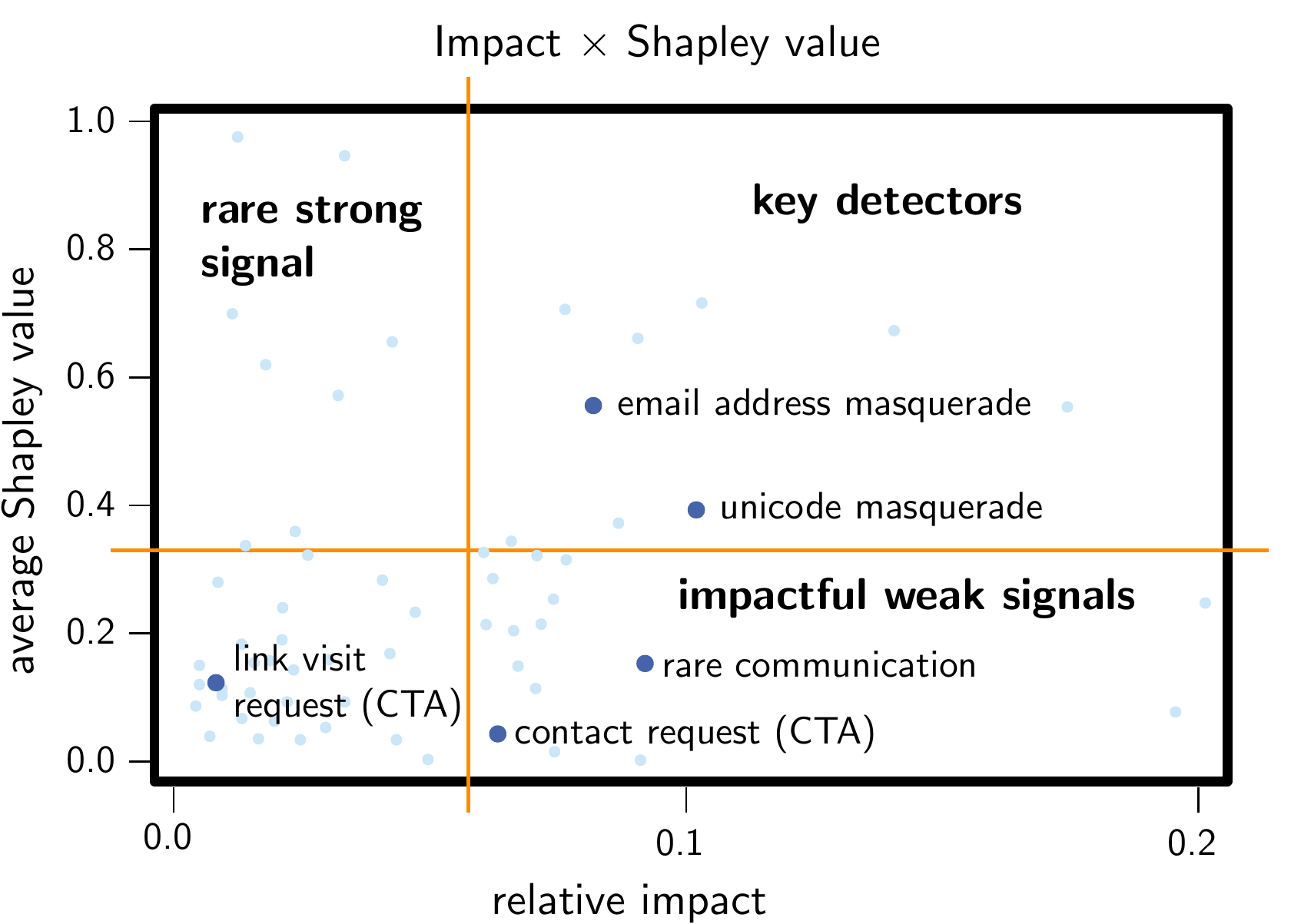}
\caption{Detector performance measured by average Shapley values and relative impact. A subset of detectors are highlighted in different quadrants of the space. The quadrants can be interpreted according to the labels in the figure. A detection can have high average Shapley value, requiring less accompanying detections to convict, but if it occurs rarely then its overall impact will be limited.
} \label{fig:shapley}
\end{figure}

Second, average \emph{Shapley values}~\cite{shapley1953value,lundberg2017unified} in convicted emails, where the detector occurs, are used to measure contribution of individual detectors towards the outcome. This concept, originally from cooperative game theory, is a widely used feature importance measure.
The relative impact and average Shapley values are being tracked over time and visualised using plots similar to Figure~\ref{fig:shapley}.
Both measures provide key insights into the combined performance of detection and classification layers as they are affected by the frequencies of detection occurrences, correlations between detections and by the logic inside of the classification layer that combines detections into convictions.

\section{Conclusion}

We have described CAPE, a modular system for detection of BEC attacks, successfully tested in an enterprise production environment for over two years.
By focusing on modularity and the integration of domain knowledge with machine learning approaches applied across multiple email modalities such as text or images, CAPE overcomes practical challenges posed by computational limitations, unavailability of representative datasets, distribution drift etc.
Furthermore, CAPE provides explainable verdicts, offering valuable insights for users such as SOC analysts in their investigations.

We have discussed what constitutes a BEC attack, the various design considerations behind CAPE and presented its architecture as well as a selected subset of detectors.
Further, we detailed a continuous efficacy monitoring process and a Bayesian framework for integration of customer feedback which both are essential to improve efficacy and adapt to changes in the threat landscape on an ongoing basis.

A potential limitation of this study is the absence of predictive performance measurement using an established public dataset.
To the best of our knowledge, there is no available public dataset that includes a representative sample of BEC emails and effectively captures the extensive complexity of benign traffic.
Datasets without well-represented benign emails would prove inadequate, as they would produce skewed and overly optimistic results.
The development of such a dataset is hindered by the extreme sensitivity and subsequent restrictive access policies concerning the benign portion of the traffic.
Furthermore, CAPE's usage of state, namely the learned email's context, would further complicate such evaluation.

\bibliographystyle{ieeetr}
\bibliography{bibliography}

\end{document}